\documentclass[12pt,preprint]{aastex}

\def\eq#1{equation (\ref{#1})}

\def\ap{a}
\def\au{~{\rm AU}}
\def\da{\Delta A}

\def\dat{\da_{tot}}
\def\days{~{\rm days}}
\def\deltap{\delta_{\rm p}}
\def\dos{D_{\rm os}}
\def\dls{D_{\rm ls}}
\def\dol{D_{\rm ol}}
\def\drel{D}
\def\ep{\epsilon_{\rm p}}
\def\kms{{\rm km~s^{-1}}}
\def\kpc{{\rm kpc}}
\def\mp{M_{\rm p}}
\def\ms{M_{\rm sat}}
\def\rhop{\rho_{\rm p}}
\def\rin{R_{in}}
\def\rjup{R_{Jup}}
\def\rout{R_{out}}
\def\rsch{R_{\rm Sch}}
\def\rsun{R_{\odot}}
\def\rp{R_{\rm p}}
\def\te{t_{\rm E}}
\def\thetae{\theta_{\rm E}}
\def\yr{{\rm ~yr}}

\begin{document}

\title{Probing Structures of Distant Extrasolar Planets With Microlensing}

\author{B.\ Scott Gaudi\footnote{Hubble Fellow}}
\affil{School of Natural Sciences, Institute for Advanced Study, 
       Princeton, NJ 08540, U.S.A.}
\email{gaudi@sns.ias.edu}

\author{Heon-Young Chang}
\affil{Korea Institute for Advanced Study,
207-43 Cheongryangri-dong Dongdaemun-gu, Seoul 130-012, Korea}
\email{hyc@ns.kias.re.kr}

\author{Cheongho Han}
\affil{Department of Physics, Institute for Basic Science Research, 
       Chungbuk National University, Chongju 361-763, Korea}
\email{cheongho@astroph.chungbuk.ac.kr}

\begin{abstract}

Planetary companions to the source stars of a caustic-crossing binary
microlensing events can be detected via the deviation from the parent
light curves created when the caustic magnifies the star light reflecting
off the atmosphere or surface of the planets.  The magnitude of the
deviation is $\deltap \sim \ep \rhop^{-1/2}$, where $\ep$ is the
fraction of starlight reflected by the planet and $\rhop$ is the
angular radius of the planet in units of angular Einstein ring radius.
Due to the extraordinarily high resolution achieved during the caustic
crossing, the detailed shapes of these perturbations are sensitive to 
fine structures on and around the planets.  We consider the 
signatures of rings, satellites, and atmospheric features on  
caustic-crossing microlensing light curves.  We find that, for 
reasonable assumptions, rings produce deviations of order $10\% \deltap$, 
whereas satellites, spots, and zonal bands produce deviations of order 
$1\%\deltap$.  We consider the detectability of these features using 
current and future telescopes, and find that, with very large apertures 
($>$30m), ring systems may be detectable, whereas spots, satellites, 
and zonal bands will generally be difficult to detect.  We also present 
a short discussion of the stability of rings around close-in planets, 
noting that rings are likely to be lost to Poynting-Robertson drag on 
a timescale of order $10^5$ years, unless they are composed of large 
($\gg$1~cm) particles, or are stabilized by satellites.

\end{abstract}
\keywords{gravitational lensing -- planets: rings -- planets and 
	  satellites: general}

\section{Introduction}

Precise radial velocity surveys have detected over 100 planetary
companions to FGKM dwarf stars in the solar neighborhood (see
http://cfa-www.harvard.edu/planets/catalog for a list of planets and
discovery references).  Among the interesting trends that have been uncovered 
in this sample of planets 
are a positive correlation between the frequency of planets and 
metallicity of the host 
stars \citep{gonzalez1997, gonzalez1998, laughlin2000, santos2001,
reid2002}, a paucity of massive, close-in planets
\citep{zucker2002,patzold2002}, and a `piling-up' of less-massive, close-in planets
near periods of $P\simeq 3\days$.  This latter trend is important
because the number of planets which transit their parent stars is
roughly proportional to $1/a$, where $a$ is the semi-major axis.  
The discovery and interpretations of these global trends 
provide clues to the physical mechanisms
that affect planetary formation, migration, and survival.  

A somewhat different way of obtaining clues about the physical
processes at work in planetary systems is to acquire detailed
information about individual planets.  With radial velocity
measurements alone, such information is limited only to the minimum
mass $\mp\sin{i}$ of the planet, and the semi-major axis $a$ and
eccentricity of its orbit.  However, if the planet also transits its
parent star, then it is possible to infer considerably more
information.  A basic transit measurement allows one to infer the
radius, mass, and density of the planet, as has been done with the
only known transiting extrasolar planet, HD209458b
\citep{charbonneau00, henry2000}.  This in turn allows one to place
constraints on the planet's orbital migration history
\citep{burrows2000}.  More detailed photometric and spectroscopic data
during (and outside of) the transit can be used to study the
composition of, and physical processes in, the planetary atmosphere
\citep{seager2000, seager2000b, charbonneau2002, brown2002}, measure
the oblateness, and thus constrain the rotation rate, of the planet
\citep{hui2002, seager2002}, and to search for rings and satellites
associated with the planet \citep{sartoretti99, schneider99, brown01}.

The `classical' method of searching for planets via microlensing was
first proposed by \citet{mao91}, and subsequently further developed by
\citet{gould92}.  In this method, a planetary companion to the primary 
lens star produces a small perturbation atop the smooth, symmetric
lensing light curve created by the primary.  The microlensing
method has several important advantages over other methods, as well as 
several disadvantages (see \citealt{gaudi2003} for a review).  The most
important advantage is that the strength of the planet's signal
depends weakly on the planet/primary mass ratio and thus it is the
only currently feasible method to detect Earth-mass planets
\citep{bennett96}.  The other advantage is that it enables one to
detect planets located at large distances of up to several tens of
kiloparsecs.  However, it also has disadvantages, the most important
of which is that the only useful information one can obtain is the
mass ratio between the planet and the primary.  Thus classical
microlensing searches only allow one to identify the existence of the
planet, and build statistics about the types of planetary systems, but
cannot be used to obtain detailed information about the discovered
planets.  This is especially problematic in light of the fact that
follow-up of the discovered systems will generally be difficult or
impossible.

Recently, \citet{graff00} and \citet{lewis00} proposed a novel method
of detecting planets via microlensing.  They suggested that one could
detect close-in giant planets orbiting the {\it source} stars of
caustic-crossing binary-lens events via accurate and detailed photometry 
of the binary-lens light curve.  In this method, the planet can be 
detected because the light from the planet is sufficiently magnified 
during the caustic crossing to produce a noticeable deviation to the 
lensing light curve of the primary.  The magnitude of the deviation is 
$\deltap \sim \ep \rhop^{-1/2}$, where $\ep$ is the ratio of the 
(unlensed) flux from the planet to the (unlensed) flux from the star, 
and $\rhop$ is the angular radius of the planet in units of the angular 
Einstein ring radius $\thetae$ of the lens system.  The Einstein ring 
radius is related to the physical parameters of the lens system by 
\begin{equation}
\thetae=\sqrt{{{2 \rsch}\over \drel} },
\label{eqn:thetae}
\end{equation}
where $\rsch=2GM/c^2$ is the Schwarzschild radius of the lens, $M$ 
is the total mass of the lens, $\drel\equiv \dos\dol/\dls$, and 
$\dos$, $\dol$, and $\dls$ are the distances between the observer-source, 
observer-lens, and lens-source, respectively.  For searches in the 
optical, the light from the planet will be dominated by the reflected 
light from the star, and $\ep \sim 10^{-4}$ for close-in 
planets.\footnote{Because the fraction of reflected light decreases 
as $\ep \propto a^{-2}$, optical searches will generally only be 
sensitive to close-in planets.  However, planets may have significant 
intrinsic flux in the infrared, enabling the detection of more distant 
companions at longer wavelengths.}   Adopting typical parameters, 
$\rho_{\rm p}\sim 10^{-4}$, and thus $\deltap \sim 1\%$.  \citet{graff00}
demonstrated that this level of photometric precision is currently
within reach of the largest aperture telescopes.  The exquisite 
resolution afforded by caustics may allow one to study features
on and around the source in detail, and with larger aperture telescopes, 
one may able to study spots and bands on the surfaces of detected planets 
by looking for small deviations to the nominal light curve \citep{graff00}.  
Here we study the signatures of these and other structures on lensing 
light curves, quantify the magnitude of the deviations, and assess 
their detectability using current and future instrumentation.  
Specifically, we consider the signatures of rings and satellites, as 
well as atmospheric features such as spots, zonal bands, and scattering.
\citet{lewis00} considered using variations in the
polarization during the planetary caustic crossing to 
probe the composition of the planetary atmosphere.   
The effects of the phase of the planet on the light curve were considered 
previously by \citet{ashton01}.

The layout of the paper is as follows.  In \S\ref{sec:binlens}, we
discuss binary lenses and their associated caustic structures, and
describe the magnification patterns near caustics.  We discuss
expectations for the existence, stability, and properties of rings,
satellites, and atmospheric features of close-in extrasolar planets in
\S\ref{sec:gen}.  In \S\ref{sec:quantitative}, we layout the formalism
for calculating microlensing light curves, 
and apply this formalism to make a quantitative predictions 
for the deviations caused by planets (\S\ref{sec:planet}), satellites 
(\S\ref{sec:satellites}), rings (\S\ref{sec:rings}), and atmospheric 
features (\S\ref{sec:atmosphere}). We address the detectability of 
these deviations in \S\ref{sec:detect}, and summarize and conclude 
in \S\ref{sec:summary}.

\section{Binary Lenses and Caustic Crossings\label{sec:binlens}}

\begin{figure*}
\epsscale{1.0}
\centerline{\plotone{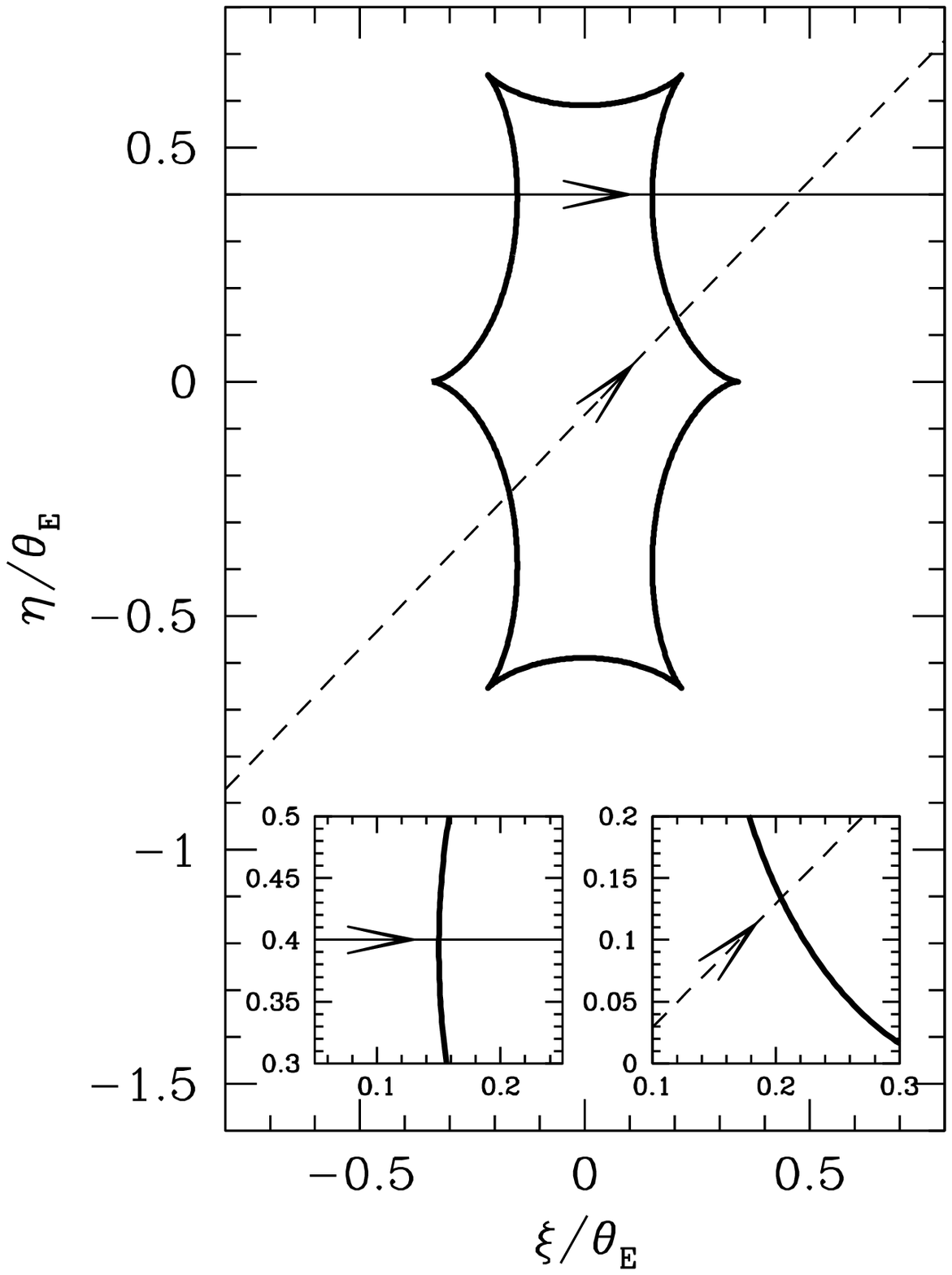}}
\caption{
An example of the caustic structure (thick solid curve) produced by a 
binary lens.  In this case, the system has equal mass components 
separated by one Einstein ring radius.  The coordinates are centered at 
the midpoint of the binary, and all lengths are normalized by the Einstein 
ring radius.  The solid and dashed straight lines are the two source 
trajectories considered in the numerical simulations. The insets are the 
details of the regions around the second caustic crossings of these 
trajectories.
}
\label{fig:caustic}
\end{figure*}

If a microlensing event is caused by a lens system composed of two
masses, the resulting light curve can differ dramatically from
the symmetric curve due to a single lens event.  The main new feature
of binary lens systems is the formation of caustics.  Caustics are the
set of positions in the source plane $(\xi, \eta)$ on which the
magnification of a point source is formally infinite.  The set
of caustics form closed curves, which are composed of multiple concave
line segments that meet at points.  The concave segments are referred
to as fold caustics, whereas the points are cusps. The number and
shape of caustic curves varies depending on the separation and the
mass ratio between the two lens components.  Figure \ref{fig:caustic} 
shows an example caustic structure of a binary lens system with equal 
mass components separated by $\thetae$.  For more details on the caustic 
structure of binary lenses, see \citet{schneider86} and \citet{erdl93}.  
The caustic cross section generally decreases with decreasing mass 
ratio, and decreases for widely and closely separated components.  
Therefore, the majority of caustic-crossing binary-lens events will 
have caustic structures similar to that shown in Figure \ref{fig:caustic}.  
Most source trajectories (straight lines through the source plane) 
will not intersect the caustic near cusps, therefore the majority of 
caustic crossings will be simple fold caustic crossings.  Near a 
fold caustic, the total magnification $A$ of a point source is 
generically given by \citep{schneider92, gaudi2002},
\begin{equation}
A(u_\perp) = \left(\frac{u_\perp}{ u_r} \right)^{-1/2} + A_0,
\label{eqn:afold}
\end{equation} 
where $u_\perp$ is the angular normal distance of the source from the
fold in units of $\thetae$, $u_r$ is related to the local derivatives
of the lens potential, and describes the effective `strength' of the
caustic (also in units of $\thetae$), and $A_0$ is the magnification 
of all the images unrelated to the caustic.  The divergent nature of 
the magnification (as $A\propto u_\perp^{-1/2}$) translates into high 
angular resolution in the source plane.

Due to the divergent magnification near a caustic, the light
curve of an caustic-crossing binary lens event is characterized by
sharp spikes which are generally easily detectable.  \citet{mao91}
predicted that $\sim 7$\% of all events seen toward the Galactic
bulge should be caustic-crossing events.  Analyses of the
databases of the MACHO \citep{alcock00} and OGLE \citep{jaroszynski02}
lensing surveys have demonstrated that binary microlensing events are being
detected at a rate roughly consistent with theoretical predictions.
Therefore a significant sample of caustic-crossing binary-lens events
should be available each year for planet searches.  The planetary caustic 
crossing is expected to occur within $\sim 1~{\rm day}(a/0.1\au)$
of the stellar caustic crossing.  Therefore, in order to detect close-in 
planets and their associated structures, extremely dense sampling for 
a period of $\sim 2$ days centered around one of the stellar caustic 
crossings is required.  Caustic crossings occur in pairs, and 
although the first caustic crossing may not be detected real time 
because of its short duration, it can be inferred afterwards from the 
enhanced magnification interior to the caustic.  Followup observations 
can therefore be prepared before the second caustic crossing, and dense 
sampling throughout the second caustic crossing will be possible.

The usefulness of caustic-crossing binary-lens events has already been
demonstrated in numerous ways.  Precise photometry during the caustic
crossings of several events has been used to measure the limb
darkening profiles of stars located in both the Galactic bulge and the
Small Magellanic Cloud \citep{albrow1999,afonso00,albrow2001,an2002},
and spectra taken during the unusually long caustic crossing of one
event has been used to resolve the atmosphere of a $K$-giant in the
bulge \citep{castro2001,albrow2001b}.  It also has been proposed that
irregular structures on the source star surface such as spots can be
studied in detail by analyzing the light curves of caustic-crossing
binary lens events \citep{han00, chang02}.  The same principles that
make these measurements possible also allow one to study tiny
structures on and around the planetary companions to the source stars
of caustic-crossing events.

What is the ultimate resolution that can be obtained during a caustic
crossing?  In geometric optics, this is set by the frequency of
measurements and the number of photons that can be acquired in a given
measurement, so that extremely small structures in the source plane
could, in principle, be probed with sufficiently large telescopes and
sufficiently high cadence.  However, for very small sources, geometric
optics breaks down, and diffraction effects become important.  For a
fold caustic crossing, this occurs when the angular size $\rho_{\rm s}$ 
of the source in units of $\thetae$ satisfies $\rho_{\rm s}< \rho_d$, 
where \citep{ulmer95,jaroszynski95}, 
\begin{equation}
\rho_d \equiv \left( \frac{\lambda}{16\pi\rsch}\right)^{2/3}u_r^{-1/3}.
\label{eqn:rhod}
\end{equation}
Here $\lambda$ is the wavelength of the light.  For $\lambda\simeq 
800~{\rm nm}$ ($I$-band) and $M=0.3M_\odot$, $\rho_d \simeq 7\times 
10^{-8} u_r^{-1/3}$.  For nearly equal-mass binaries, such as shown in 
Figure \ref{fig:caustic}, $u_r\sim 1$.  Thus the ultimate resolution 
achievable during a caustic-crossing is $\rho_d\thetae\sim 20~{\rm pas}$, 
or a length of $\rho_d\thetae\dos\sim 26~{\rm km}$ at the distance of 
the bulge.   This is considerably smaller than the size of any of the 
structures we consider here, so we can safely ignore diffraction effects.

\section{Satellites, Rings, and Atmospheric Features of Close-in Planets\label{sec:gen}}

All of the planets in our solar system have at least one satellite,
with the exceptions of Mercury and Venus, so it seems at least
plausible that satellites are common by-products of the formation of
planetary systems.  Satellites are perturbed by the tidal bulges they
induce on their parent planets, and their orbits evolve under the
influence of this torque.  If the timescale for this evolution is
sufficiently short, the satellite will either spiral inward until
impacting with the parent planet, or outward until it reaches the Hill
radius of the planet, and is lost to the parent star.  The
survival of satellites has been considered by numerous authors in the
context of our solar system (see, e.g.\ \citealt{ward1973}).  More
recently, \citet{barnes02} studied the lifetimes of satellites in the
context of extrasolar planetary systems, and found that satellites
with $\ms > M_\oplus$ cannot survive for more than $\sim 5$~Gyr around
Jupiter-like planets with separations $a\la 0.25\au$, assuming a
solar-mass primary star. Therefore relatively massive satellites
around close-in planets are expected to be rare.  However, the
lifetime of satellites depends sensitively on the (uncertain) tidal $Q$
value of the planet, and therefore may be significantly in error.
Furthermore, planets around lower-mass stars are expected to retain
their satellites for a longer period of time.

Satellites of close-in planets are unlikely to have substantial
atmospheres because their surface escape speeds are generally small
enough that most light element gases will have evaporated over the age
of the system.  The equilibrium blackbody temperatures of satellites
of planets with $a<0.1\au$ are $T\ga 900~{\rm K}$, and any gases with
an atomic mass less than $\sim 18$ (including ${\rm
H,~He,~CH_4,~NH_3}$ and ${\rm H_2O}$), will have been entirely lost over the $\sim 5~{\rm Gyr}$ age
of the system for satellites with mass $\le M_\oplus$.  Therefore, the scattering surface of any extant
satellite will likely be rocky.

Ring systems exist around all of the gas giants in our solar system,
and therefore might also be expected to be common debris from
planetary formation.  The rings we observe in the solar system have
varied properties, but at least some, such as the rings of Saturn, are
composed of icy materials which give rise to high albedos.  Such high
albedos would aid considerably in detection in the current context.
Unfortunately ices cannot exist as separations less than,
\begin{equation}
a \simeq \left( \frac{L_*}{16 \pi \sigma
T_{sub}^4}\right)^{1/2}=2.7\au\left(\frac{L_*}{L_\odot}\right)^{1/2},
\label{eqn:snowline}
\end{equation}
where $L_*$ is luminosity of the parent star and $T_{sub}=170{\rm K}$ is 
the sublimation temperature of ice.  Thus close-in planets cannot have 
rings composed of icy material.  Rocky rings are not precluded; however 
the constituent particles are subject to numerous dynamical forces, 
including Poynting-Robertson (PR) drag, viscous drag from the planet 
exosphere, torques from satellites and/or shepherd moons, and internal 
collisions.  The characteristic decay time for PR-drag is 
\citep{goldreich82}
\begin{equation}
t_{\rm PR}\sim 10^5\yr
\left(\frac{\rho}{{\rm g~cm^{-3}}}\right)
\left(\frac{r}{{\rm cm}}\right)
\left(\frac{a}{{\rm 0.1~AU}}\right)^2,
\label{eqn:prdrag}
\end{equation}
where $\rho$ and $r$ are the density and radius of the particles,
respectively.  For close-in planets, the decay time $t_{vd}$ for
viscous drag is likely to be considerably larger than $t_{\rm PR}$
\citep{goldreich82}, unless the planet's exospheres are quite dense,
$\rho \ga 10^{-16}~{\rm g~cm^{-3}}$.
Interparticle forces only serve to spread the ring.  Therefore, the 
dominant effect, aside from satellite perturbations, is likely to be 
PR drag.  It is clear that rings of close-in planets will be lost to 
the planet on a relatively short timescale unless they are stabilized 
by interactions with satellites.  Since, as we 
have just discussed, satellites around close-in planets are themselves 
generally not long-lived, it is not clear that this is a viable method 
of maintaining rings.  A definitive exploration of the stability of 
rings and satellite systems around close-in planets is beyond the scope 
of this paper, but warrants future study.

An enormous amount of effort has been devoted to modeling of the
atmospheres of extrasolar planets, with ever increasing levels of
sophistication (see, e.g.\ \citealt{saumon1996,burrows1997,seager2000}).  
Special emphasis has been placed on close-in planets \citep{seager1998,
goukenleuque2000}.  The recent confrontation of observations of the 
radius and atmosphere of HD209458b \citep{brown01,charbonneau00} with 
theoretical predictions \citep{seager2000,burrows2000} generally 
indicates that considerably more work needs to be done \citep{guillot2002,
fortney2003}.  The problem of modeling the atmospheres of close-in
planets is an especially difficult one, due to the fact that these
planets are tidally locked to their parent stars, and subject to
strong stellar irradiation.  It seems likely that non-equilibrium
processes, weather, and photoionization will all play a role in
accurate theoretical models.  It is therefore perhaps a bit premature
to speculate on the existence and nature of surface features, such as
zonal bands and spots, on extrasolar planets.  There are intriguing 
indications, however, that close-in planets may possess large-scale 
surface features.  \citet{showman2002} and \citet{cho2002} 
showed that the extreme day-night 
temperature difference in close-in, synchronized planets drive large-scale 
zonal winds that can reach $\sim 2~\kms$.  These circulation patterns 
result in non-uniform surface temperatures which may lead to significant 
variations in the scattering, reflecting, and absorbing properties of 
the atmosphere.

Regardless of whether or not surface irregularities exist on
extrasolar planets, it is clear that a uniform surface brightness,
which has been assumed in previous microlensing simulations
\citep{graff00,lewis00,ashton01}, will likely not be an accurate
representation of the global illumination pattern of the planet.  Even
for simple Lambert scattering, in which each area element of the
planetary atmosphere reflects the incident flux uniformly back into
the $2\pi$ available solid angle, the surface brightness profile of
the planet is non-uniform due to projection effects.  Departures from 
Lambert scattering are expected, and depend on such properties as the 
particle size of the condensates in the atmosphere \citep{seager2000}.

To summarize, it remains unclear whether satellites or ring systems
can exist around close-in extrasolar planets.  Models of the
close-in planetary atmospheres have not yet reached the level of
sophistication required to definitively predict whether large-scale surface
features such as zonal bands or spots will be present.  It seems quite 
likely, however, that the surface brightness profile of extrasolar
planets will {\it not} be uniform, as has previously been assumed when
calculating the effects of microlensing.  We will therefore proceed
with rampant optimism, and assume that all of the above structures,
(rings, moons, zonal bands, spots, and non-uniform surface
brightness profiles) may exist, and consider the nature and
magnitude of their effects on microlensing light curves.

\section{Quantitative Estimates\label{sec:quantitative}}

In this section, we estimate the magnitude of the features in lensing 
light curves produced by satellites, rings, and atmospheric features, 
relative to the nominal light curve produced
by an isolated, circular planet
with uniform surface brightness.  For the most part, we use semi-analytic 
means to produce quantitative estimates in order to elucidate the 
dependence of the deviations on the parameters of the system.  However, 
in some cases the signals cannot be computed analytically.  We therefore 
complement our semi-analytic results with detailed numerical simulations.

The time-dependent total flux $F(t)$ from a system composed of a star
and $N$ additional source components being microlensed can be
generically written as,
\begin{equation}
F(t)=F_* A_*(t) + \sum_i^{N} F_{i}A_{i}(t) + B,
\label{eqn:fmaster}
\end{equation}
where $F_*$ is the unlensed flux of the star, $A_*$ is the magnification 
of the star, $F_{i}$ and $A_{i}$ are the unlensed flux and magnification 
of the $i$th additional component (which may include planets, satellites, 
rings, etc.), and $B$ is any unlensed blended flux.  We will henceforth 
assume no blending, but note that any deviation in the light curve is 
suppressed in the presence of blending by a factor $( 1 + f_b/A_{nb})^{-1}$,
where $f_b\equiv B/(F_*+\sum F_{i})$ is the ratio of blend to total
source flux, and $A_{nb}$ is the total magnification in the absence of
blending.

Generally, we have that $F_{i} \ll F_*$, and $\sum F_{i} \ll F_*$, 
and thus the observed magnification can be approximated as
\begin{equation}
A_{obs}\equiv \frac{F(t)}{F_*+\sum F_i} \simeq A_* + 
\dat;~~~~~\dat\equiv\sum_i f_{i}A_{i},
\label{eqn:aobs}
\end{equation}
where we have defined $\dat$, the extra magnification due to the 
planetary companion and associated structures, and the flux ratio 
$f_{i}\equiv F_i/F_*$ between the $i$th component and the star.

For a uniform, circular source 
sufficiently close to a linear fold caustic, the magnification is 
\citep{chang84,schneider92},
\begin{equation}
A_{fs}= \left(\frac{u_r}{\rho_{\rm s}}\right)^{1/2}G_0(z) + A_0,
\label{eqn:afs}
\end{equation}
where $z\equiv u_\perp/\rho_{\rm s}$ is the normal distance of the 
source from the caustic in units of the source size.  The
function $G_0(z)$ describes the normalized light curve for a uniform,
circular source crossing a fold caustic, and can be expressed in terms
of elliptic integrals \citep{schneider1987}.  It is useful because it
can be used to describe the magnification of any source that can
decomposed into components with azimuthal symmetry.  $G_0$ has a
maximum of $\simeq 1.38$ at $z\simeq-0.66$.  In the range $-1\le z \le
1$, the mean value of $G_0(z)$ is $0.95$ and the RMS is $1.04$.  The
magnification can also be written as a function of time by defining
$t_r\equiv u_r \te \csc\gamma$, and $z=(t-t_{cc})/\Delta t$, where
$t_{cc}$ is the time when the center of the source crosses the
caustic, and $\gamma$ is the angle of the trajectory with respect to
the caustic, and the timescale of the caustic crossing is $\Delta
t\equiv\rho_{\rm s}\te\csc\gamma$.  Here $\te = \thetae/\mu$ is the
timescale of the primary event, and $\mu$ is the relative lens-source
proper motion.

For typical microlensing events toward the Galactic bulge, 
$\thetae\simeq 320\mu{\rm as}$ and $\mu\simeq 12.5~{\rm km~s^{-1} 
{kpc^{-1}}}$, and thus $\te\simeq 22\days$.  We will assume that the 
primary source is a G-dwarf in the bulge, i.e.\ that it has a radius 
$R_*=\rsun$ and is located at the distance $\dos=8~\kpc$.  The angular 
radius is $\theta_*\simeq 0.6\mu{\rm as}$, and thus the dimensionless 
source size is $\rho_*=\theta_\ast/\theta_{\rm E}= 1.8\times 10^{-3}$.  
The caustic crossing timescale for a source of physical radius $R$ is,
\begin{equation}
\Delta t_\ast \simeq 1~{\rm hr}~\csc{\gamma} \left( \frac{R}{\rsun}\right).
\label{eqn:deltat}
\end{equation}
Thus the primary caustic crossing is expected to last 
$2\Delta t_*\sim 2~{\rm hr}$ for $R_*=R_\odot$.

\subsection{Planet\label{sec:planet}}

The largest contribution to $\dat$ generally will be from the planet itself, 
$\da_{\rm p}\equiv f_{\rm p}A_{\rm p}$.  In the case of light reflected 
by a planetary atmosphere or surface, the flux fraction $f_{\rm p}$ will 
depend on the radius $\rp$ of the planet, its distance $\ap$ from the 
star, the scattering properties of the atmosphere, and the phase of the 
planet.  Generically, the flux ratio $f_{\rm p}$ between the planet and star 
can be written as \citep{sobolev75},
\begin{equation}
f_{\rm p}=\epsilon_{\rm p} \Phi(\alpha),
\label{eqn:fluxrat}
\end{equation}
where $\alpha$ is the phase angle, defined as the angle between the 
star and Earth as seen from the planet, $\Phi(\alpha)$ is the phase function, 
and $\epsilon_{\rm p}$ is the flux ratio at $\alpha=0$,
\begin{equation}
\epsilon_{\rm p}={\cal A}_{\rm p} \left(\frac{\rp}{\ap}\right)^2\simeq 2.28 \times 10^{-5} 
{\cal A}_{\rm p}\left(\frac{a}{0.1\au}\right)^{-2} \left(\frac{\rp}{\rjup}\right)^{2}
\label{eqn:epsilon}
\end{equation}
Here ${\cal A}_{\rm p}$ is the geometric albedo of the planet.  For a Lambert sphere, ${\cal A}_{\rm p}=2/3$, 
and
\begin{equation}
\Phi(\alpha)=\frac{1}{\pi}\left[\sin{\alpha} + (\pi-\alpha)\cos{\alpha}\right].
\label{eqn:lambertphase}
\end{equation}

The magnification $A_{\rm p}$ of the planet will depend on the size
of the planet, as well as on its surface brightness.  The surface
brightness, in turn, depends on the phase of the planet, as well as the 
scattering properties of the atmosphere.  The effects of the phase of the planet on 
the magnification have been considered by \citet{ashton01}, and we 
consider the effects of Lambert-sphere scattering on the magnification
in \S\ref{sec:lambert}.   We will therefore assume that the planet has 
a uniform surface brightness and is at full phase ($\alpha=0$ and thus 
$\Phi=1$), unless otherwise stated. In this case, the 
magnification of the planet is simply given by \eq{eqn:afs}, with 
$\rho_{\rm s}=\rho_{\rm p}$, where $\rho_{\rm p}$ is the angular
radius of the planet in units of $\thetae$.  Adopting this expression, 
the contribution of the perturbation from the planet is,
\begin{equation}
\da_{\rm p} = \epsilon_{\rm p} \left({u_r \over \rho_{\rm p}}\right)^{1/2} 
G_0(z_{\rm p}).
\label{eqn:dap}
\end{equation}
Note that we have assumed that $f_{\rm p} A_0 \ll 1$.  In most cases, 
$A_0$ is of order unity, and this will be an excellent approximation.  
We will furthermore assume that $f_i A_0 \ll 1$ in deriving all analytic 
expressions.  During the planetary caustic crossing, the RMS\footnote{The 
RMS is the relevant quantity for signal-to-noise considerations, see 
\S\ref{sec:detect}.} of $G_0$ is $\sim 1$.  Thus the magnitude of the 
deviation is given by the coefficients of the $G_0$ function in \eq{eqn:dap}, 
\begin{equation}
\delta_{\rm p} \equiv \epsilon_{\rm p} \left({u_r \over \rho_{\rm p}}
\right)^{1/2}.
\label{eqn:deltap}
\end{equation}
For binary lenses with caustic structures similar to that shown in Figure 
\ref{fig:caustic}, $u_r$ is of order unity \citep{lee1998}.

Thus for typical microlensing
 bulge parameters, $\delta_{\rm p}=1.67\times 10^{-3}{\cal A}_{\rm p} (\rp/\rjup)^{3/2}
(\ap/0.1\au)^{-2}$, and for a planet with properties similar to HD209458b 
($\rp=1.347\rjup$ and $\ap=0.0468\au$, \citealt{brown01}), and ${\cal A}_{\rm p}=2/3$, 
$\delta_{\rm p}\sim 0.8\%$.  From \eq{eqn:deltat}, the duration of planetary
caustic crossing is $2 \Delta t_{\rm p} \simeq 16~{\rm min}~\csc\gamma$ for $\rp=1.347\rjup$.

\subsection{Satellites\label{sec:satellites}}

The magnitude of the deviation caused by a satellite can be estimated using 
the same formalism as used for the planet (\S\ref{sec:planet}).  As for the 
planet, we will assume that the satellite has a uniform surface brightness 
and phase $\alpha=0$.  The deviation $\da_{\rm sa}$ caused by the satellite
is then,
\begin{equation}
\da_{\rm sa} = \epsilon_{\rm sa} \left({u_r \over \rho_{\rm sa}}\right)^{1/2} 
G_0(z_{\rm sa}),
\label{eqn:das}
\end{equation}
where $\epsilon_{\rm sa}$ is the flux ratio between the satellite and star, 
and $\rho_{\rm sa}$ is the dimensionless size of the satellite.  In analogy 
to the case of the planet alone, we can define $\delta_{\rm sa}\equiv
\epsilon_{\rm sa} ({u_r/\rho_{\rm sa}})^{1/2}$ to be the magnitude of the 
deviation from the satellite.  If we assume that the distance of the satellite 
from the planet is small compared to $\ap$, we can relate this to the magnitude 
of the deviation due to the planet by,
\begin{equation}
\delta_{\rm sa} = \kappa_{\rm sa} \left( \frac{R_{\rm sa}}{\rp}\right)^{3/2} 
\delta_{\rm p},
\label{eqn:deltas}
\end{equation}
where $\kappa_{\rm sa}={\cal A}_{\rm sa}/{\cal A}_{\rm p}$ is the ratio of the 
geometric albedos of the satellite and planet, and $R_{\rm sa}$ is the radius 
of the satellite.  The ratio $\kappa_{\rm sa}$ will depend quite strongly 
on the compositions of the planet and satellite.  
 As we
argued in \S\ref{sec:gen}, satellites of close-in planets are unlikely
to have substantial atmospheres, and therefore the scattering surface
will be rocky in composition.
Albedos of rocky bodies depend on their composition, but are generally low, 
$\la 0.3$.   For definiteness, we will assume that $\kappa_{\rm sa}=0.1$,
and assume a satellite radius of $R_{\rm sa}=0.2\rp$.  This is $\sim 2R_\oplus$ 
for a Jupiter-size planet.  Then $\delta_{\rm sa} \simeq 1\%\delta_{\rm p}$. 

\begin{figure*}
\epsscale{0.8}
\centerline{\plotone{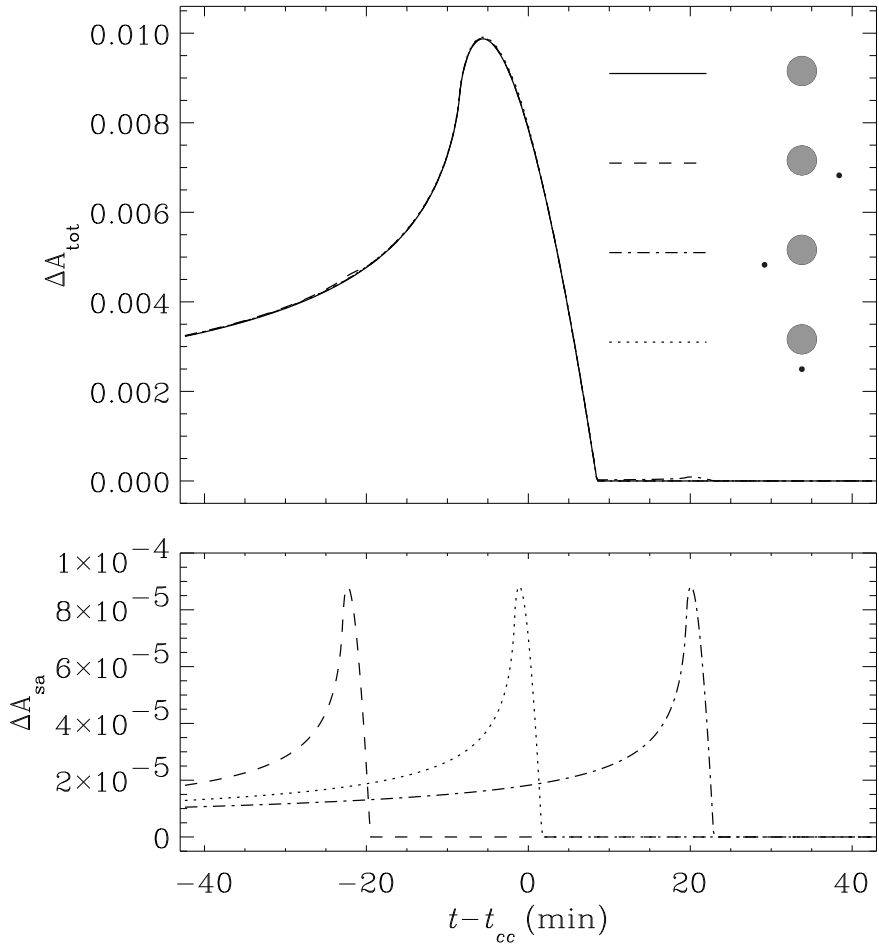}}
\caption{
Top: Caustic-crossing microlensing light curves from a planet with and 
without a satellite with radius equal to 20\% of the radius of the planet,
and albedo equal to 10\% of the planet's albedo.  The solid line shows 
the light curve from the planet only.
The other lines show the magnification
including a satellite with various relative positions. 
The events arise from the second caustic crossing of the dashed source trajectory
through the binary-lens systems depicted in Fig.\ \ref{fig:caustic}.
Bottom: The additional magnification $\Delta A_{\rm sa}$ caused by the 
satellite alone.  
}
\label{fig:sat}
\end{figure*}

Assuming typical bulge parameters, an analog of the HD209458 system at
$\dos=8\kpc$, and for the second caustic crossing of the dashed
trajectory in Figure~\ref{fig:caustic}, which has the properties
$\gamma=71^\circ$ and $u_r=0.82$, we find $\delta_{\rm p}=7.19\times
10^{-3}$ and $\Delta t_{\rm p}=8.48~{\rm min}$.  Adopting these parameters,
Figure \ref{fig:sat} shows the total magnification associated with the
planet and satellite, $\dat=\da_{\rm p} + \da_{\rm sa}$, as well as the
extra magnification from the satellite alone, $\da_{\rm sa}$.  We
show the effect for various satellite positions.  The time it takes
for the caustic to cross the satellite is $2\Delta t_{\rm sa}\simeq
3.4~{\rm min}$.

\subsection{Rings\label{sec:rings}}

Rings of extrasolar planets have such low mass that they have no 
observable dynamical effect on the host star's motion.  However, as 
shown by the example of Saturn's ring, they can be significantly 
more extended than planets.  This makes them much easier
to be identified by transits, for which the signal varies
as the area $\Omega$ of the feature that occults the star,
and microlensing, for which the signal 
is $\propto \Omega^{3/4}$ due to the competing effects
of the amount of reflected light ($\propto \Omega$), and the magnification
($\propto \Omega^{-1/4}$).

We can obtain an analytic estimate of the signal caused by a ring
by assuming a face-on geometry.  We model the ring as a circular annulus 
with uniform surface brightness, outer radius $R_{out}$, and inner radius 
$\rin$.    The magnification from the ring is then,
\begin{equation}
\da_{\rm ri} = \delta_{\rm ri} H,
\label{eqn:dari}
\end{equation}
where we have defined
\begin{equation}
H\equiv \left[ \left(\frac{\rout}{\rp}\right)^{3/2}G_0(z_{out})-
\left(\frac{\rin}{\rp}\right)^{3/2}G_0(z_{in})\right],
\label{eqn:hfunc}
\end{equation}
and  $z_{out}\equiv z_{\rm p} (\rp/\rout)$, and similarly for $z_{in}$.
The shape and magnitude of the function $H$ will depend on the relative 
sizes of $\rout$, $\rin$, and $\rp$, but will generally be of order unity 
for the ring systems considered here.  The magnitude of the deviation due 
to the ring is therefore roughly $\delta_{\rm ri}$, which in terms of 
the deviation from the planet is, 
\begin{equation}
\delta_{\rm ri}=\kappa_{\rm ri} \delta_{\rm p},
\label{eqn:deltari}
\end{equation}
where $\kappa_{\rm ri}={\cal A}_{\rm ri}/{\cal A}_{\rm p}$ is the ratio 
of the geometric albedos of the ring and planet.  We argued in 
\S\ref{sec:gen} that any surviving ring systems of close-in planets must 
be rocky in nature, and thus the albedos will generally be small.  Adopting 
$\kappa_{\rm ri}=0.15$, $\delta_{\rm ri}=15\% \delta_{\rm p}$.

\begin{figure*}
\epsscale{0.8}
\centerline{\plotone{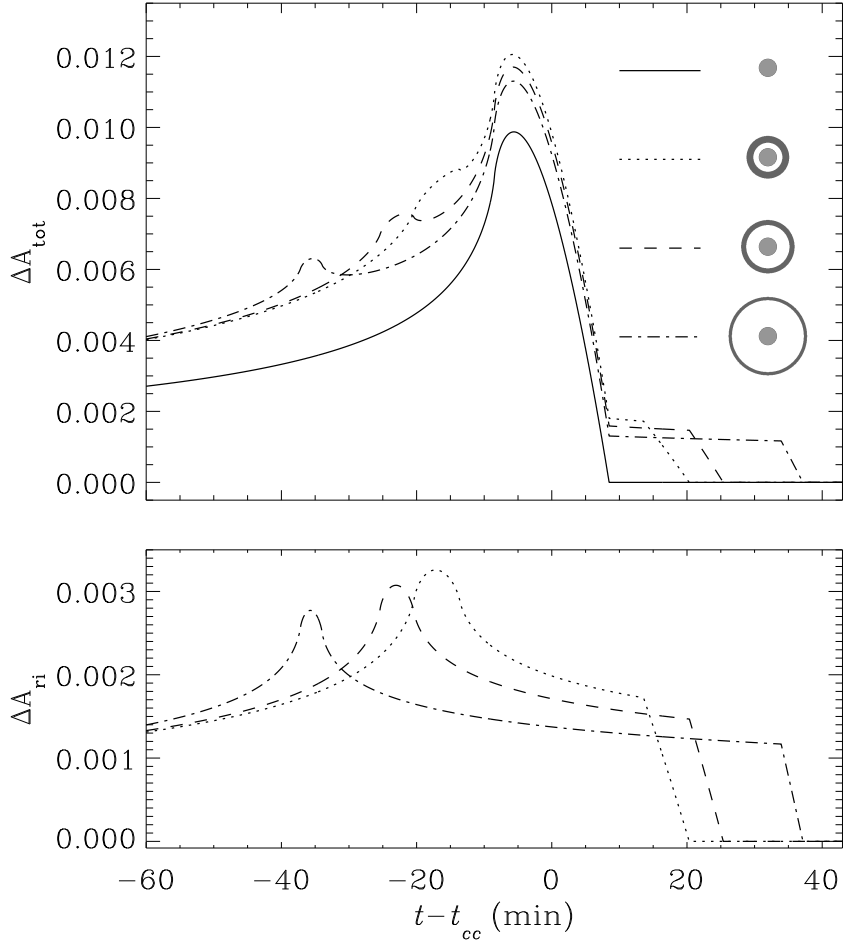}}
\caption{
Top: Caustic-crossing microlensing light curves from a planet with and
without a face-on ring with different gaps between the planet's disk 
and the ring.  The solid line shows the light curve from the planet only. 
The icons show the ring geometries for 
the corresponding light curves with rings.  The inner ring radius is 
$R_{\rm ri}=1.6$ (dotted), 2.4 (short-dashed), and $4.0\rp$ (long-dashed), 
respectively, where $\rp$ is the radius of the planet.  The outer ring 
radius is adjusted so that the projected area of the ring is equal to 
$\pi (R_{out}^2-R_{in}^2)=10\rp^2$.  The ring has an albedo equal
to 15\% of the albedo of the planet.  The lens system and the 
source trajectory responsible for the events are the same as for the 
light curves presented in Fig.\ \ref{fig:sat}.
Bottom: The additional magnification $\Delta A_{\rm ri}$ caused by the ring
alone. 
}
\label{fig:ring1}
\end{figure*}

Figure \ref{fig:ring1} shows the total magnification associated with
the planet and ring system, $\dat=\da_{\rm p} + \da_{\rm ri}$, where we
have adopted the same parameters for the planetary
deviation as in \S\ref{sec:satellites}, namely
 $\delta_{\rm p}=7.19\times 10^{-3}$ and $\Delta t_{\rm p}=8.48~{\rm min}$. 
We also show the magnification 
from the ring system alone, $\da_{\rm ri}$.
We show 
the effect of the ring for various inner radii, and in order to isolate 
the effect of varying the inner radius on the resulting light curve, we 
fix the total area of the ring systems to be $\pi(\rout^2-\rin^2)=10\rp^2$.  
The two small bumps on the left and right sides of the primary peak are 
caused by the ring's entrance and exit of the caustic.  The time
between the two ring-induced bumps (or between one of the bumps and
the primary peak), relative to the time scale of the planetary
perturbation, is a measure of the relative dimension
of the ring compared to the size of the planet disk.  The time it
takes for the caustic to cross the ring system is $2\Delta t_{\rm ri}=2
(\rout/\rp)\Delta t_{\rm p}$.  For the largest ring system shown in Figure
\ref{fig:ring1}, this is $\sim 74~{\rm min}$.  We find that the signal
of the ring generally decreases slowly as the gap between the planet
and the ring increases.

Although our analytical expressions for the deviations induced by a
simple, face-on ring system are useful in that they allow one to gain
insight and find relatively simply scalings for the magnitude of the
effect, they are limited in their scope.  In particular, they cannot
be used to access the effects of inclined ring systems.  For this,
numerical integration must be employed in order to calculate the
magnification of the ring.  We consider sources crossing the 
second caustic of the dashed
trajectory in Figure~\ref{fig:caustic}.  To calculate the light curve,
we first compute the full binary-lens magnification on each area
element on the surface of the source and then average the
magnifications of the individual elements, weighting by the surface
brightness of each element.  We again assume typical microlensing parameters,
and a source system analogous to HD 209458 at 8 kpc.  This yields
$\epsilon_{\rm p}=1.26\times 10^{-4}$, and $\rho_{\rm p}\te=8.02~{\rm min}$.  The
effects of $u_r$ and $\gamma$ are included implicitly in our numerical
integration of the binary-lens magnification for the specific trajectory
we have adopted.  For
the planet-only case, we find our numerical and analytic light curves
agree quite well, indicating that our numerical integrations are accurate,
the caustic is well-approximated by a simple linear fold, and that
we are using the appropriate values of $\gamma$ and $u_r$ in our analytic
expressions. 

In the numerical simulations, we model ringed planets in the same
manner as the analytic calculations.  We assume that the ring is an 
infinitely thin annulus without any gap, with inner and outer radii 
$R_{in}$ and $R_{out}$, respectively.  We assume the planet has 
a uniform albedo of ${\cal A}_{\rm p}=2/3$
for pure Lambert scattering.  We assume that the ring
has an albedo of ${\cal A}_{\rm ri}=0.1$, and thus $\kappa_{\rm ri}=0.15$,
unless otherwise specified.  
For the simulations, we take the effects of the planet's phase and the 
ring's inclination into consideration.  For some specific geometries of 
the planet, host star, and the observer, the planet will cast a shadow 
on the ring.  We also take this effect into consideration.  We then 
investigate the variations of the pattern of ring-induced anomalies 
depending on these various factors affecting the shape of the ring.

Defining the projected shape of a ringed planet requires many
parameters, such as the inclination of the ring, the phase angle, the
radius of the planet disk, and the inner and outer ring radii.  As a
result of the large number of parameters, it is often difficult to 
imagine the planet's shape based on these parameter values.  We therefore 
simply use small icons to characterize the planet and ring shapes instead 
of specifying all these parameters whenever we present light curves 
resulting from specific realizations.

\begin{figure*}
\epsscale{0.8}
\centerline{\plotone{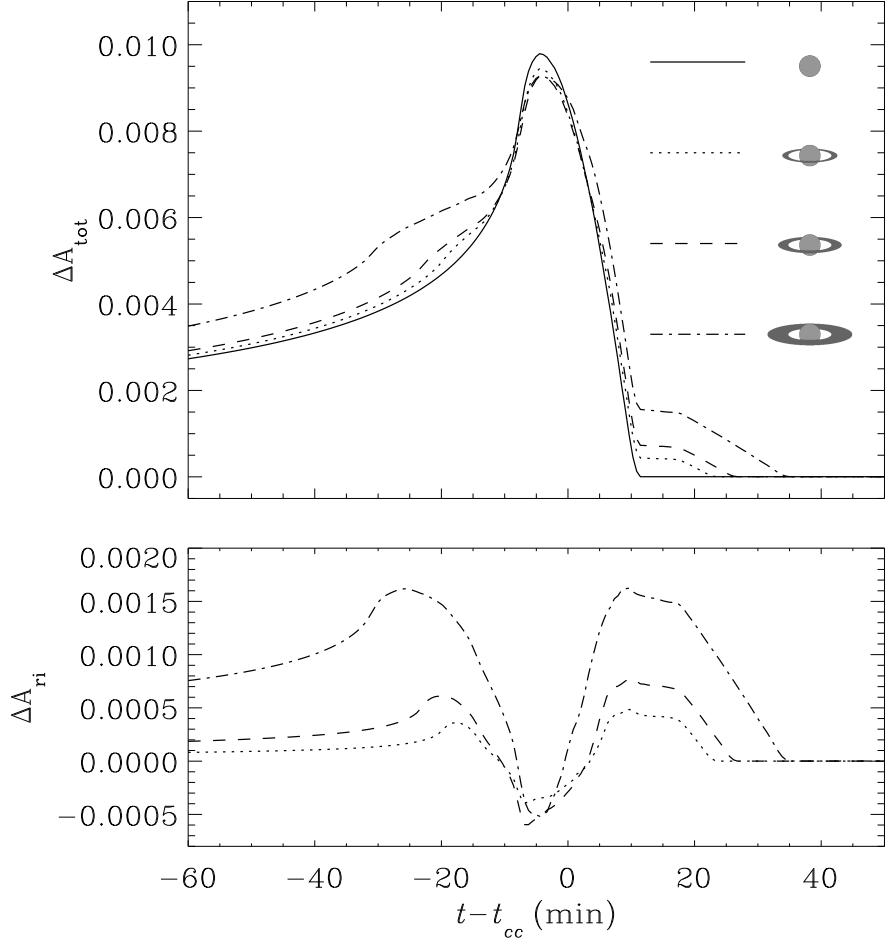}}
\caption{
Top: Caustic-crossing microlensing light curves from a planet with inclined 
rings of different widths.  The rings have a common inner radius of 
$R_{in}=2.0\ \rp$, but different outer ring radii of $R_{out}=2.6$, 
3.0, and 4.0 $\rp$.  The ring has an inclination of $i=75^\circ$, and
albedo relative to the planet of $15\%$.  
The lens system and the 
source trajectory responsible for the events are the same as for the 
light curves presented in Fig.\ \ref{fig:sat}.  Bottom: The additional 
magnification $\Delta A_{\rm ri}$ caused by the ring alone. 
}
\label{fig:ring2}
\end{figure*}

Figure \ref{fig:ring2} shows the effect of the width of the ring on
the light curve.  All of the systems have a common inner ring radius
and gap between the planet disk and the ring, but different outer ring
radii.  We have also assumed that the ring system is viewed at an
angle of $i=75^\circ$ with respect to the normal to the ring plane.
Not surprisingly, increasing the width of the ring increases the
magnitude and duration of the ring-induced perturbation.

\begin{figure*}
\epsscale{0.8}
\centerline{\plotone{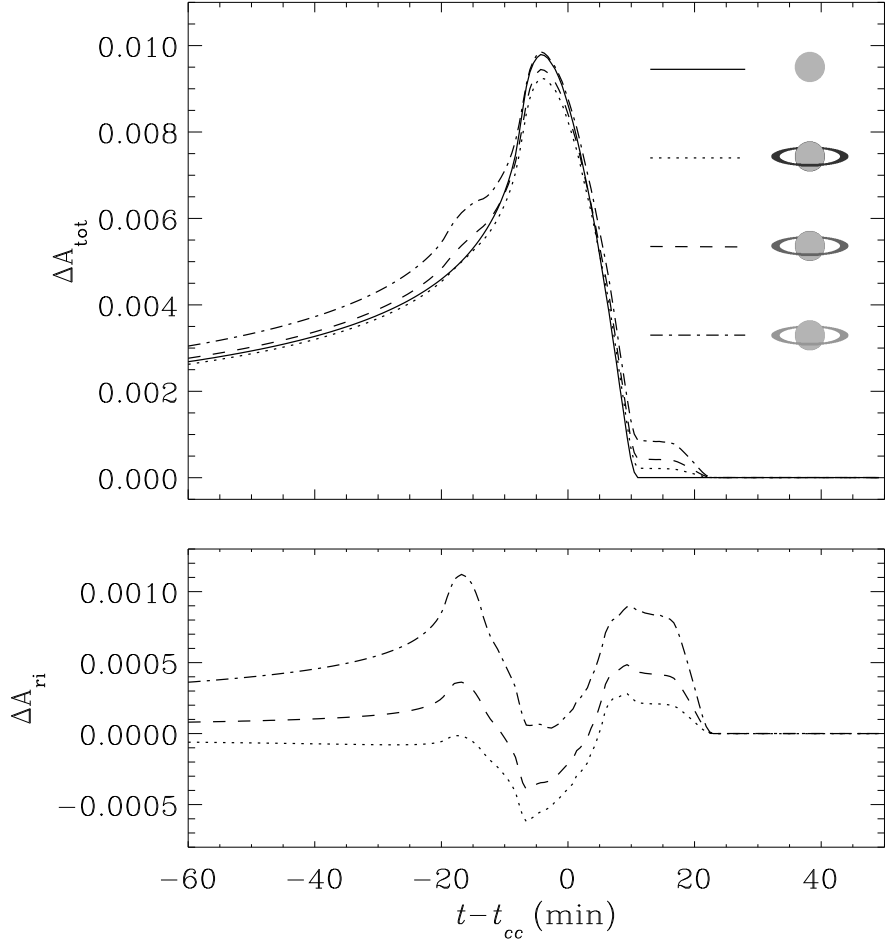}}
\caption{
Top:  Caustic-crossing microlensing light curves from a planet with inclined 
rings with different albedos.  The total magnification is shown for a 
planet with rings of equal size, [inner and outer ring radii of 
$(R_{in}, R_{out})= (2.0, 2.6)\rp$], but relative albedos of 
$\kappa_{\rm ri}=0.075$ (solid), 0.3 (dotted), and 0.6 (dashed).  The 
inclination of the rings is $i=75^\circ$.  The lens system and the 
source trajectory responsible for the events are the same as for the 
light curves presented in Fig.\ \ref{fig:sat}.  Bottom: The additional 
magnification $\Delta A_{\rm ri}$ caused by the ring alone. 
}
\label{fig:ring3}
\end{figure*}

Figure \ref{fig:ring3} shows the dependence of the ring-induced
anomaly pattern on the albedo of the ring particles.  We test three
different albedos of ${\cal A}_{\rm ri}=0.05$, 0.1, and 0.2 
(corresponding to $\kappa_{\rm ri}=0.075$, 0.3, and 0.6), and the 
difference in the greyscale of the rings in the icons represents 
the variation of the albedo.  As expected, the magnitude of the ring 
signal is proportional to the albedo of ring particles.

\begin{figure*}
\epsscale{0.8}
\centerline{\plotone{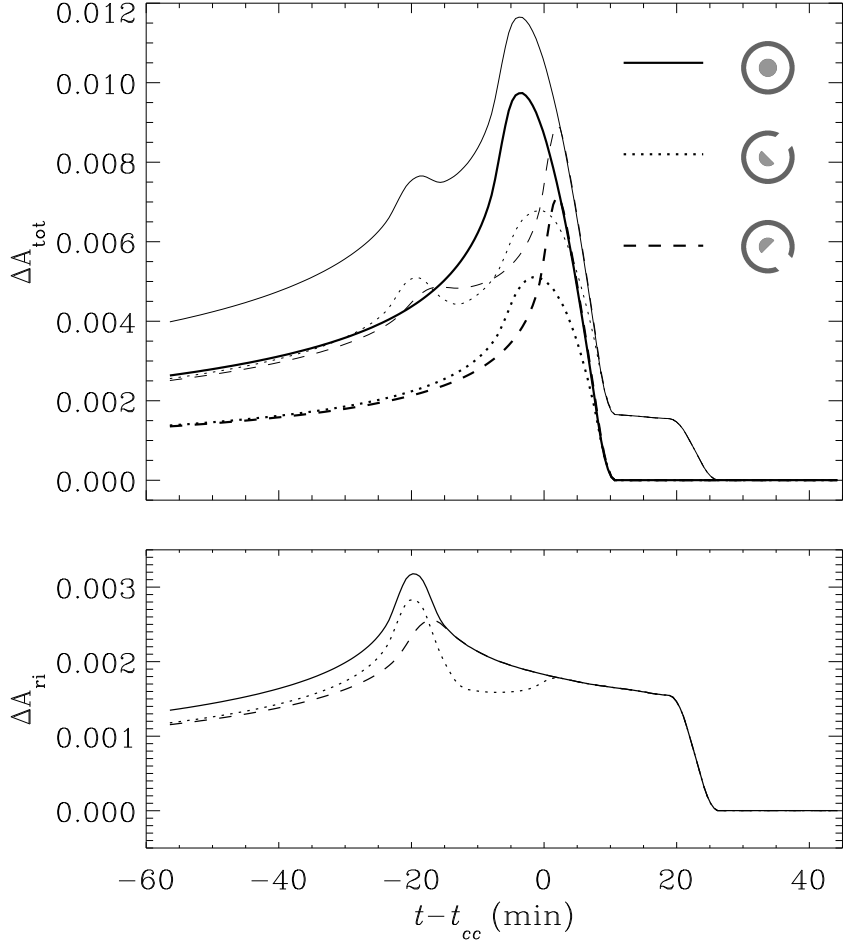}}
\caption{
Top: Caustic-crossing microlensing light curves from a planet with a ring, 
where the planet's shadow is cast on the ring.  Note that, due to the 
specific geometry of the planet, host star, and observer, the planet is 
at quarter phase.  Each ring has inner and outer ring radii of 
$(R_{in}, R_{out})=(2.4, 3.0)\rp$, is seen face-on, and has an albedo relative
to the planet of 15\%. The lens 
system and the source trajectory responsible for the events are the same 
as for the light curves presented in Fig.\ \ref{fig:sat}.  For all line types,
the heavier curve is for the planet only, while the lighter curve is for the planet
and the ring.  Bottom:  The additional 
magnification $\Delta A_{\rm ri}$ caused by the ring alone. 
}
\label{fig:ring4}
\end{figure*}

In Figure \ref{fig:ring4}, we demonstrate the effects of a shadow cast
on the ring by the planet.  In this case, we must also take into
account the phase of the planet to be self-consistent.  For the
geometry depicted in Figure \ref{fig:ring4}, the planet is at quarter
phase.  We find that the shape of the part of the light curve arising
from the planet is strongly dependent on the phase of the planet, as
discussed in detail by \citet{ashton01}, however the shape of the
signal due the ring does not depend strongly on the effect of the shadow
of the planet, due to the relatively small surface area of the ring
occulted by the planet.

From the light curves in Figures \ref{fig:ring2} -- \ref{fig:ring4},
we conclude that the typical magnitude of ring-induced deviations is
$\delta_{\rm ri}\sim \kappa_{\rm ri}\deltap$, confirming our analytic estimates,
although there are considerable variations depending on the
inclination and size of the ring.

\subsection{Atmospheric Features\label{sec:atmosphere}}

In contrast to the signatures of satellites and rings, atmospheric
features that are located on the surface of the planet will only
produce detectable deviations while the planet is resolved.  The
planet is only effectively resolved when it is within about one
planet radius from the caustic.
Therefore, deviations caused by spots, bands, or otherwise non-uniform
surface brightness profiles will only be noticeable during a time $\sim
2\Delta t_{\rm p}$ centered on the caustic crossing; outside of this the flux
from the planet will essentially be given by
the unresolved flux (i.e.\ the mean surface
brightness times the area of the planet disk) times the magnification
of a point-source at the center of the planet disk.

\subsubsection{Spots\label{sec:spots}}

Spots, such as the Great Red Spot on Jupiter, have been observed on
the gas giants in our solar system, and are regions of cyclonic
activity that have slightly different temperatures and pressures from
their surrounding atmospheres.  As a result, the colors and albedos of
spots are also slightly different.  These spots can be quite large 
relative to the planetary radii; the Great Red Spot has a size larger 
than $R_\oplus$.  In this subsection, we provide an analytic estimate of 
the effect of a spot on the microlensing light curve.  We model the 
spot as a circle of radius $R_{\rm sp}$, with an albedo equal to a 
fraction $\kappa_{\rm sp}$ of the albedo of the remainder of
the planetary surface.  The deviation caused by the spot is then,
\begin{equation}
\da_{\rm sp}=
\delta_{\rm p}
\left\{ 
\frac{1-\kappa_{\rm sp}}{1-(1-\kappa_{\rm sp}) 
\left({R_{\rm sp}}/{\rp}\right)^{2}} 
\left[
\left(\frac{R_{\rm sp}}{\rp}\right)^{2}G_0(z_{\rm p})-
\left(\frac{R_{\rm sp}}{\rp}\right)^{3/2}G_0(z_{\rm sp})
\right]
\right\}.
\label{eqn:dasp}
\end{equation}
Note that, in deriving \eq{eqn:dasp},  the mean surface brightness of 
the spotted planet has been normalized to that of the planet without 
the spot.  This ensures that the magnifications of the two cases are 
identical when the source is not resolved.  That this is true can be seen 
by noting that, for $z_{\rm sp},z_{\rm p}\rightarrow -\infty$, the term 
in square brackets goes to zero, because $G_0(z_{\rm sp}) \rightarrow (R_{\rm sp}/\rp)^{1/2}G_0(z_{\rm p})$.   Since we are 
generally concerned with small spots with  $R_{\rm sp}/R_{\rm p} \ll 1$, 
we can make an estimate of the magnitude $\delta_{\rm sp}$ of the 
deviation caused by the spot by ignoring terms in \eq{eqn:dasp} of order 
$(R_{\rm sp}/R_{\rm p})^{2}$ or higher, and looking at the resulting 
coefficient to the $G_0(z_{\rm sp})$ function.  We find that
\begin{equation}
\delta_{\rm sp} \simeq -(1-r_{\rm sp})
\left(\frac{R_{\rm sp}}{\rp}\right)^{3/2}\delta_{\rm p}.
\label{eqn:deltasp}
\end{equation}
For $\kappa_{\rm sp}=0.8$, 
and $R_{\rm sp}=0.2 \rp$ ($\sim 2R_\oplus$ 
for $\rp=\rjup$),  we find $\delta_{\rm sp}\simeq 2\%\delta_{\rm p}$.

\begin{figure*}
\epsscale{0.8}
\centerline{\plotone{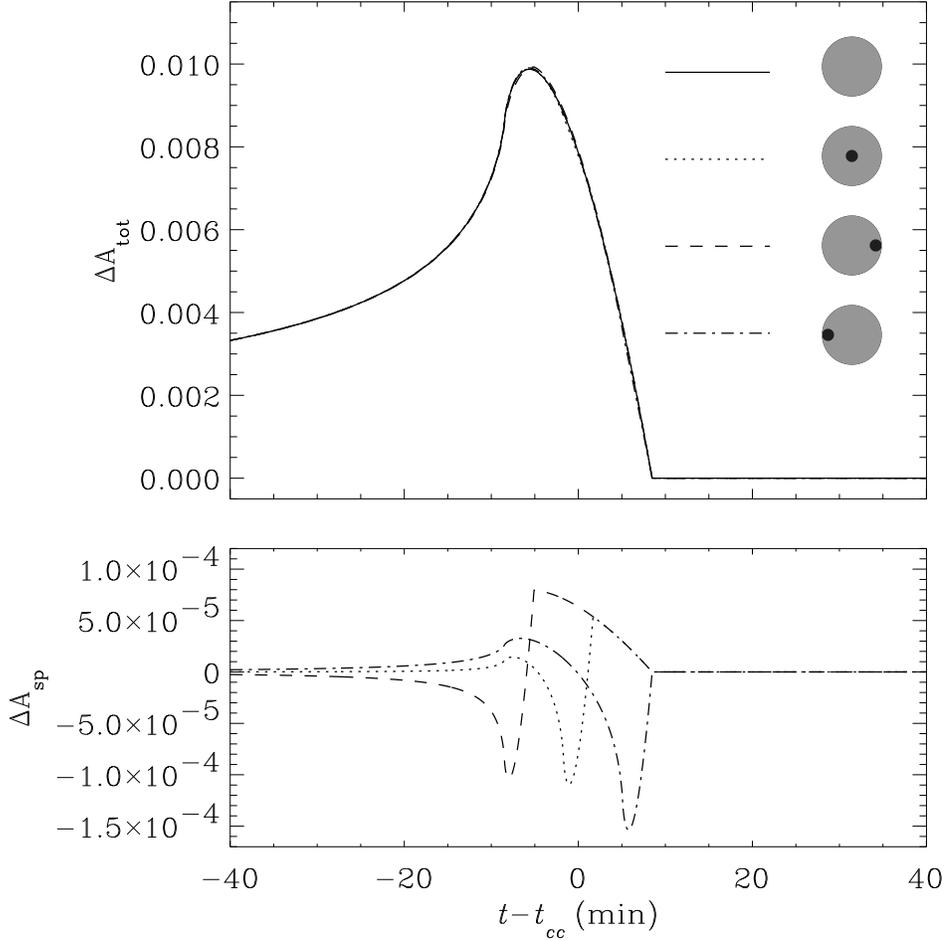}}
\caption{
Top: Caustic-crossing microlensing light curves from a planet with and 
without a spot with radius equal to 20\% of the radius of the planet,
and albedo equal to 80\% of the planet's albedo.  The solid line shows 
the light curve from the planet only,
the magnitude of the deviation due to the planet.  The other lines show 
the magnifications including a spot with various relative positions. 
The lens 
system and the source trajectory responsible for the events are the same 
as for the light curves presented in Fig.\ \ref{fig:sat}.
Bottom: The deviation $\Delta A_{\rm sp}$ due to the spot.  
}
\label{fig:spot}
\end{figure*}

In Figure \ref{fig:spot}, we show the total magnification $\dat$, with and 
without a spot of radius 
$R_{\rm sp}=0.2 \rp$ and relative albedo $\kappa_{\rm sp}=0.8$.  
We have adopted the same parameters as in \S\S~\ref{sec:satellites} and \ref{sec:rings},
$\delta_{\rm p}=7.19\times 10^{-3}$ and $\Delta t_{\rm p}=8.48~{\rm min}$.
We vary the position of the spot as shown.  We also show 
$\da_{\rm sp}$,  the deviation caused by 
the spot from the light curve of a planet with uniform surface brightness.  
We find the magnitude of the deviation caused by the spot to be 
$\sim 1\% \delta_{\rm p}$, in rough agreement with our analytic estimate.
Since, for linear fold caustics, the magnification is independent
of position along the direction parallel to the caustic,
the results in Figure \ref{fig:spot} are applicable for any circular spot 
located on the star at the same perpendicular distance from the caustic as the
spots shown.  Microlensing of spotted planets is analogous to microlensing of
spotted stars, see \citet{han00}, \citet{lewis2001}, and \citet{chang02} for 
examples of such lightcurves.

\subsubsection{Zonal Bands\label{sec:zonal}}

As can be seen on the surface of Jupiter,  
gaseous giant planets may exhibit color variations on their surface,
e.g.\ zonal bands, which will cause surface brightness variations within
a given spectral band.  In this subsection, we investigate whether 
 zonal bands can produce noticeable signatures in 
lensing light curves.

We model zonal bands by stripes which are parallel with the equator of 
the planet. We assume that the albedos of these stripes alternate with 
relative values $\kappa_{\rm ba}$, and we vary the total number of 
stripes.  Analytic estimates are generally impossible for arbitrary 
inclinations of the planet, however, for the special case when the 
planet is seen pole-in, the pattern of the zonal bands is simply 
concentric annuli with alternating albedos.  In this case, we can find 
a semi-analytic expression for the deviation from a uniform surface
brightness due to the zonal bands.  The resulting expression is 
somewhat complicated,
\begin{equation}
\da_{\rm ba}=
\delta_{\rm p}
\left[ 
\frac{1-\kappa_{\rm ba}}{1-(1-\kappa_{\rm ba}) 
{(\tilde R_{\rm ba}/R_{\rm p})^2}}
H' \right],
\label{eqn:dab}
\end{equation}
where 
\begin{equation}
H' \equiv 
\left(\frac{\tilde R_{\rm ba}}{\rp}\right)^{2}G_0(z_{\rm p})-
\sum_{i=1}^{N_{\rm ba}-1}(-1)^{i+\ell}\left(\frac{R_{{\rm ba},i}}
{\rp}\right)^{3/2}G_0(z_{{\rm ba},i}),
\end{equation}
and $N_{\rm ba}$ is the number of bands, $R_{{\rm ba},i}$ is 
the outer radius of the $i$th band, $z_{b,i}\equiv z_{\rm p} 
(\rp/R_{{\rm ba},i})$, $\ell=1$ if $N_{\rm ba}$ is odd, and 
$\ell=0$ if $N_{\rm ba}$ is even, and
\begin{equation}
{\tilde R_{\rm ba}^2} \equiv \sum_{i=1}^{N_{\rm ba}-1}
(-1)^{i+\ell}R_{{\rm ba},i}^{2}.
\label{eqn:normalization}
\end{equation}
Note that, as $N_{\rm ba}\rightarrow \infty$, $\da_{\rm ba}\rightarrow 0$. 
The term in square brackets in \eq{eqn:dab}
cannot be reduced to a simpler expression,
and must be calculated explicitly using the analytic
form for the $G_0$ function.

\begin{figure*}
\epsscale{0.8}
\centerline{\plotone{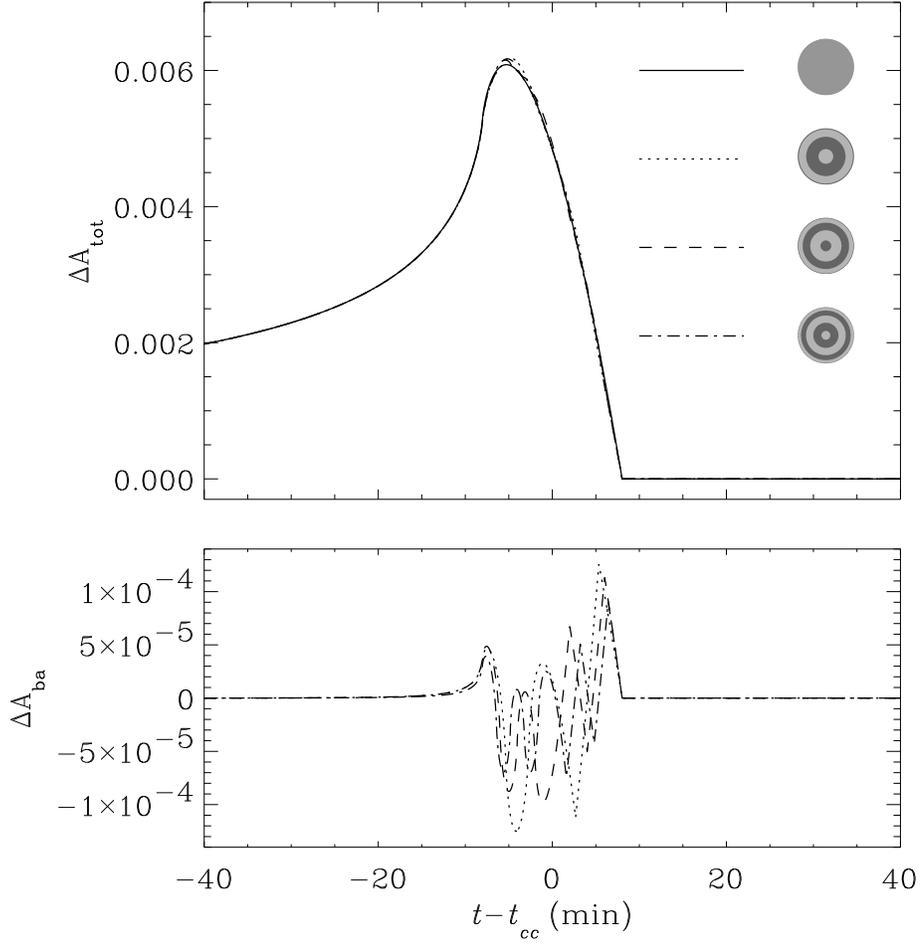}}
\caption{
Top: Caustic-crossing microlensing light curves from a planet with and 
without surface albedo variations caused by zonal bands.  The zonal bands  
are parallel with the planet's equator and cause alternate variations 
of the surface brightness, and the planet is seen pole-on.  The albedos 
of the bright and dark regions differ by $20\%$, and the mean surface 
brightness of each model has been adjusted in order to match that from 
the uniform planet.   The events arise from the second caustic crossing of the solid source trajectory
through the binary-lens systems depicted in Fig.\ \ref{fig:caustic}.
Bottom: The deviation $\Delta A_{\rm ba}$ caused 
by the zonal bands. 
}
\label{fig:band1}
\end{figure*}

In 
Figure \ref{fig:band1}, we show the total magnification $\dat$ of the 
planet, with and without $N_{\rm ba}$ 
bands with relative albedos $\kappa_{\rm ba}=0.8$, similar to that of 
the Jupiter \citep{pilcher71}.  We have adopted the same parameters as in 
the previous sections, except here we consider the second caustic
crossing of the solid trajectory in 
Figure~\ref{fig:caustic}, which has the properties $u_r=0.31$ and $\gamma=89^\circ$.
This yields $\delta_{\rm p}=4.41\times 10^{-3}$ and $\Delta t_{\rm p}=8.02~{\rm min}$. 
We vary the number of bands from 
$N_{\rm ba}=3-5$, and the bands have equally-spaced radii (and thus 
cover different surface areas on the source).  We also show 
$\da_{\rm ba}$,  the normalized deviation caused 
by the banded structure from the light curve of a planet with uniform 
surface brightness.   We find the magnitude of the deviation caused
by the bands to be 
\begin{equation}
\delta_{\rm ba}\sim 30\% (1-\kappa_{\rm ba})N_{\rm ba}^{-\beta}
\delta_{\rm p}.
\label{eqn:deltab}
\end{equation}
Here the scaling with $\kappa_{\rm ba}$ is only approximate.  The 
scaling with $N_{\rm ba}$ depends on the geometry of the zones; 
$\beta=1$ for zones with equally-spaced $R_{{\rm ba},i}$ (as shown 
in Fig.\ \ref{fig:band1}), whereas $\beta=1/2$ for equal-area bands.  
For equal-area zones, the numerical coefficient in \eq{eqn:deltab} is 
also somewhat smaller, $\sim 20\%$.

\begin{figure*}
\epsscale{0.8}
\centerline{\plotone{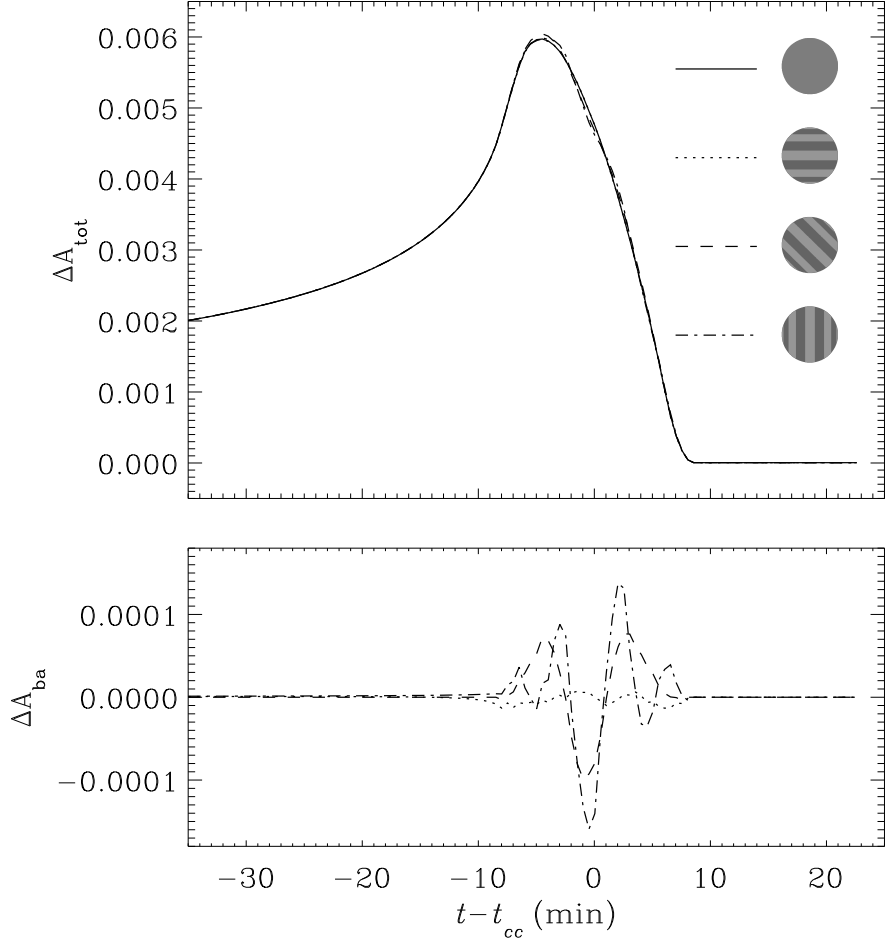}}
\caption{
Top: Caustic-crossing microlensing light curves from a planet with and without 
zonal bands.  The zonal bands are parallel with the planet's equator and 
cause alternate variations of the surface brightness, and the inclination 
of the planet is $90^\circ$.  The albedos of the bright and dark regions 
differ by $20\%$.  The solid curve is the lensing light curve corresponding 
to a planet with uniform surface brightness.  
The lens 
system and the source trajectory responsible for the events are the same 
as for the light curves presented in Fig.\ \ref{fig:band1}.
 Bottom:  The deviation $\Delta A_{\rm ba}$ caused 
by the zonal bands. 
}
\label{fig:band2}
\end{figure*}

For geometries where the planet is not pole-on, we must resort to 
numerical calculations.  For these calculations, we assume a total 
of nine bands (with four dark lanes), with relative albedos of 
$\kappa_{\rm ba}=0.8$, as in the previous example.  The albedos are 
normalized such that the average albedo is $2/3$.
Figure \ref{fig:band2} 
shows the effect of zonal bands for a planet with inclination $i=90$ 
(i.e.  the axis of rotation in the plane of the sky), and various 
orientations of the equator with respect to the caustic. 
As before, we assume the   source trajectory indicated by 
the solid line in Figure \ref{fig:caustic}.  The solid 
curve is the light curve resulting from a planet having a uniform 
surface brightness with ${\cal A}_{\rm p}$.  From the figure, one finds that the deviations
induced by the zonal bands are $\sim 1\%\delta_{\rm p}$, similar to 
the pole-on case.  Note that these deviations are generally an order 
of magnitude smaller than the typical deviations induced by rings.

\subsubsection{Lambert Sphere Scattering\label{sec:lambert}}

\begin{figure*}
\epsscale{0.8}
\centerline{\plotone{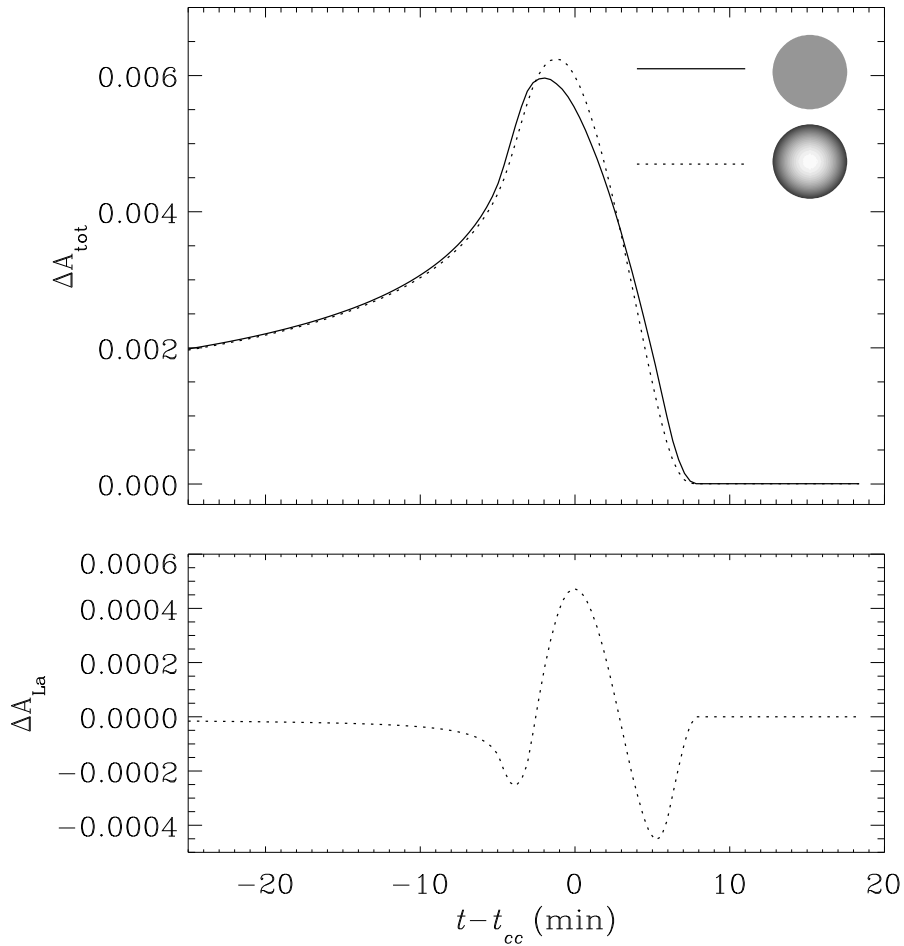}}
\caption{
Top:  Caustic-crossing microlensing lensing light curves from a planet with 
the surface brightness profile expected from Lambert scattering (dotted),
and a uniform surface brightness profile (solid).  
The lens 
system and the source trajectory responsible for the events are the same 
as for the light curves presented in Fig.\ \ref{fig:band1}.
  Bottom: The deviation $\Delta A_{\rm La}$ in the magnification
of a Lambert sphere from a source with uniform surface brightness. 
}
\label{fig:lambert}
\end{figure*}

Even without any irregular structure, the surface brightness profile
of a planet will generally not be uniform due to projection effects, 
and the scattering properties of the atmosphere. The surface brightness 
distribution will therefore vary depending on the latitude $\psi$ 
and longitude $\omega$ of the planet's surface as well as the 
planet's phase angle, $\alpha$.  To illustrate the pattern of lensing light 
curve deviations caused by a realistic atmosphere, we adopt the 
simple assumption of pure Lambert scattering, where the incident 
radiation from the host star is scattered isotropically.  Under 
this assumption, it can be shown that the surface brightness 
profile is
\begin{equation}
\label{eqn:lambertsbp}
S(\alpha,\omega,\psi) = {\bar S} \frac{3}{2}\frac{\cos(\alpha-\omega)
\cos\omega\cos^2{\psi}}{\Phi(\alpha)},\qquad {\bar S}=\frac{F_{\rm p}}{\pi\theta_{\rm p}^2}, 
\end{equation}
for $\omega\ge\alpha-\pi/2$ and zero otherwise \citep{sobolev75}. 
Here ${\bar S}$ is the mean surface brightness of the planet, and
$\Phi(\alpha)$ is the phase integral introduced in 
\S\ref{sec:planet} and displayed explicitly in \eq{eqn:lambertphase}. 
We show the surface brightness distribution for a Lambert sphere 
with $\alpha=0$ in Figure \ref{fig:lambert}, along with the resulting 
microlensing light curve.  As in \S\ref{sec:zonal}, we
have assumed the solid source trajectory in Figure \ref{fig:caustic}.  
We compare this to the light curve resulting 
from a planet with a uniform surface brightness and an albedo equal to 
the geometric albedo of the Lambert sphere (i.e.\ ${\cal A}_{\rm p}=2/3$).  We 
find that the light curve of a Lambert sphere differs from a uniform 
surface brightness profile by $\sim10\% \delta_{\rm p}$.  The light curve 
from a Lambert sphere is more highly magnified, due to the fact that 
the surface brightness profile is more centrally concentrated, and 
therefore the source is effectively smaller.  The precise shape of 
the light curve from the planet will depend on the scattering 
properties of the atmosphere, which in turn depend on the constituents 
of the atmosphere, such as the size of the condensates \citep{seager2000}.
Therefore resolution of the planetary caustic crossing would provide 
invaluable information about the physical processes in the planetary 
atmosphere.  This is complementary to the suggestion
by \citet{lewis00} of probing the planetary atmosphere 
via polizarization monitoring during the planetary caustic crossing.

\section{Detectability\label{sec:detect}}

In this section, we review the magnitudes of the effects of the
features we have considered, and assess their detectability with
current and/or future telescopes.  Table \ref{tbl:table1} summarizes
our analytic expressions for the magnitudes $\delta_x$ of the deviations 
caused by each structure $x$ that we considered (satellites, rings, 
spots, and zonal bands), in terms of the magnitude $\delta_{\rm p}$ of 
the deviation due to only the planet.  Also shown is the characteristic
timescale $\Delta t_x$ of each deviation, relative to the timescale of
the planetary caustic crossing $\Delta t_{\rm p}$.

By approximating the perturbations from each structure $x$ as boxcars 
with amplitudes $\delta_x$ and durations $\Delta t_x$,  we can write 
down {\it approximate} expressions for the ratio of the signal-to-noise 
$Q_x$ for a given perturbation to the signal-to-noise of the planetary 
pertrubation, 
\begin{equation}
\frac{Q_x}{Q_{\rm p}} \sim \left|\frac{\delta_x}{\delta_{\rm p}}\right|
\left(\frac{\Delta t_x}{\Delta t_{\rm p}}\right)^{1/2}.
\label{eqn:relativesnr}
\end{equation}
These ratios are displayed in Table \ref{tbl:table1}; they allow one 
to easily estimate the detectability of the various features in terms 
of the detectability of the planetary deviation.  For example, if one 
were to require a signal-to-noise ratio of $Q_{\rm p} = 20$ for a 
secure detection of the planet signal, then signal-to-noise of the 
deviation from a ring with relative albedo $\kappa_{\rm ri}=0.15$ 
and relative (outer) radius of $R_{\rm ri}/\rp=4$ would be 
$Q_{\rm ri}\simeq 0.15\times \sqrt{4} \times 20 = 6$.  For reasonable
parameters, we expect that $Q_x/Q_{\rm p} \ll 1$ for satellites, spots, 
and bands, whereas for rings $Q_{\rm ri}/Q_{\rm p} \sim 30\%$.

We now provide a more accurate estimate of the expected signal-to-noise 
for the various features.  Assume that a microlensing light curve is 
monitored continuously from $t_{min}$ to $t_{max}$, for a total duration 
$T$, with a telescope that collects $n_\nu$ photons per second per unit 
flux.   The signal-to-noise ratio $Q_x$ of a deviation $\Delta A_x(t)$ 
is then,
\begin{equation}
Q_x= (n_\nu T)^{1/2} { F_* \over (B+F_*)^{1/2}}\left\{ {1 \over T}  
\int_{t_{min}}^{t_{max}} dt \ [\Delta A_x(t)]^2 \right\}^{1/2},
\label{eqn:stn}
\end{equation} 
where $F_*$ is the unlensed flux of the primary star, and $B$ is the 
background flux (sky + unresolved stars).  The term in curly brackets 
is essentially the RMS of the deviation during the time of the 
observations.   We assume that the source system is an analog of 
HD20458 at $\dos=8$ kpc.  The primary has $I_*=19.5$ (a G0V star at 
$8$ kpc with 1.2 magnitudes of extinction), and its planet has 
$\rp=1.347\rjup$, $a=0.0468\au$, and ${\cal A}_{\rm p}=2/3$.  Adopting typical
bulge parameters ($M=0.3M_\odot, \dol=6~\kpc, \dos=8~\kpc$), 
and a caustic crossing with properties $u_r=1$ and $\gamma=90^\circ$, this gives 
$\delta_{\rm p}=7.93\times 10^{-3}$ 
and $\Delta t_{\rm p}=8~{\rm min}$.  We assume a total 
background flux of $19.3$, which includes the moon-averaged sky 
background at an average site, and the contribution expected from
unresolved stars in the bulge for a seeing of $0.75''$.   
We assume that a telescope of diameter $A_T$ collects 
$n_\nu=2700(A_T/10~{\rm m})^2$ photons per second at $I=20$, which 
corresponds to an overall throughput (including detector
efficiency) of $\sim 50\%$.  Finally, we 
assume that the light curve is monitored from $-5\Delta t_{\rm p}$ 
before the caustic crossing until $5\Delta t_{\rm p}$ after the 
caustic crossing of the primary star, 
for a total duration of $T\simeq 80~$min.

The resulting signal-to-noise values for the various deviations
are tabulated in Table \ref{tbl:table1}.  For 10m-class telescopes,
the deviation from the planet should be detectable with $Q_{\rm p}\simeq 15$.  
This is in rough agreement with the results of \citet{graff00} and \citet{ashton01}
for full phases.  However, \citet{ashton01} find that the signal-to-noise
depends quite strongly on the phase of the planet.  Adopting a different
phase would therefore affect the relative signal-to-noise $Q_x/Q_{\rm p}$ between
the planet and 
the satellite, ring, spot or band features, but generally not the absolute signal-to-noise $Q_x$.  
For the deviations from the other structures, we have adopted the 
parameter values appropriate to the short-dashed line in Figure 
\ref{fig:sat} for the satellite, the dotted line in Figure 
\ref{fig:ring1} for the ring, the short-dashed line in Figure 
\ref{fig:spot} for the spot, and the long-dashed line in Figure 
\ref{fig:band1} for the zonal bands.   For these perturbations we 
find signal-to-noise ratios of $Q_{\rm sa}=0.1$ (satellite), 
$Q_{\rm ri}=6.1$ (ring), $Q_{\rm sp}=0.1$ (spot), and 
$Q_{\rm ba}=0.1$ (zonal bands).  These are in rough agreement with 
the expected scaling with $Q_{\rm p}$, and generally indicate that 
it will be impossible to detect spots, bands, and satellites with 
10m-class telescopes.  Rings are potentially detectable, but 
only under somewhat optimistic scenarios, i.e.\ large, face-on rings.   
We therefore consider the detectability with larger-aperture telescopes, 
such as the proposed 30-meter aperture California Extremely Large 
Telescope (CELT) \citep{nelson2000}, or the European Space Agency's 
proposed 100-meter aperture Overwhelmingly Large Telescope (OWL) 
\citep{dierickx2000}.  We find that rings will generally be 
detectable with reasonable signal-to-noise for $A_T\ga 30~$m, and
spots, satellites and bands are likely to be undetectable with any 
foreseeable telescope.

For the deviations caused by Lambert scattering shown in Figure 
\ref{fig:lambert}, we find that $Q \sim 1\% Q_{\rm p}$.  Therefore,
the non-uniform nature of the surface brightness may be measurable 
with 100m-class telescopes.

\begin{table*}[t]
\begin{center}
\begin{tabular}{c|ccc|cccc}
\tableline
\multicolumn{1}{c}{Feature} &
\multicolumn{1}{c}{Magnitude} &
\multicolumn{1}{c}{Timescale} &
\multicolumn{1}{c}{Relative $S/N$} &
\multicolumn{4}{c}{Absolute $S/N$}\\
\multicolumn{1}{c}{} &
\multicolumn{1}{c}{$\delta_x/\delta_{\rm p}$} &
\multicolumn{1}{c}{$\Delta t_x/\Delta t_{\rm p}$} &
\multicolumn{1}{c}{$Q_x/Q_{\rm p}$} &
\multicolumn{1}{c}{10m} &
\multicolumn{1}{c}{30m} &
\multicolumn{1}{c}{50m} &
\multicolumn{1}{c}{100m} \\
\tableline
\tableline
Planet    & $\delta_{\rm p}=\epsilon_{\rm p} \left({u_r \over \rho_{\rm p}}\right)^{1/2}$ & -- &  -- & 15.1 & 45.2 & 75.4 & 150.7  \\
\hline
Satellite & $\kappa_{\rm sa} \left( \frac{R_{\rm sa}}{\rp}\right)^{3/2} $ & $\frac{R_{\rm sa}}{\rp}$ & $\kappa_{\rm sa} \left( \frac{R_{\rm sa}}{\rp}\right)^{2} $ & 0.1 & 0.2 & 0.3 & 0.7 \\
Ring      & $\kappa_{\rm ri} $ & $\frac{R_{\rm ri}}{\rp}$ &  $\kappa_{\rm ri} \left( \frac{R_{\rm ri}}{\rp}\right)^{1/2}$  & 6.1 & 18.4 & 30.7 & 61.4 \\
Spot      & $-(1-\kappa_{\rm sp})\left(\frac{R_{\rm sp}}{\rp}\right)^{3/2}$ & $\frac{R_{\rm sp}}{\rp}$ &$(1-\kappa_{\rm sp})\left( \frac{R_{\rm sp}}{\rp}\right)^{2}$ & 0.1 & 0.3 & 0.5 & 0.9 \\
Bands      & $0.3 (1-\kappa_{\rm ba})N_{\rm ba}^{-\beta}$ & 1 & $0.3(1-\kappa_{\rm ba})N_{\rm ba}^{-\beta}$ & 0.1 & 0.3 & 0.6 & 1.1 \\
\tableline
\tableline
\end{tabular}
\end{center}
\tablenum{1} {\bf Table 1} Estimated Signal-to-Noise Ratios for 
Planetary Structures. \\
\label{tbl:table1}
\end{table*}

\section{Summary and Conclusion\label{sec:summary}}

Planetary companions to the source stars of caustic crossing
microlensing events can be detected via the brief deviation created
when the caustic transits the planet, magnifying the reflected light
from the star.  The magnitude of the planetary deviation is 
$\delta_{\rm p} \sim \epsilon_{\rm p} \rho_{\rm p}^{-1/2}$, where 
$\epsilon_{\rm p}$ is the fraction of the flux of the star that is 
reflected by the planet, and $\rho_{\rm p}$ is the angular size of 
the planet in units of the angular Einstein ring radius of the lens.  
For giant, close-in planets (similar to HD20958b), $\epsilon_{\rm p} 
\sim 10^{-4}$, and for typical events toward the Galactic bulge, 
$\rho_{\rm p} \sim 10^{-4}$.  Thus $\delta_{\rm p} \sim 1\%$,
which is accessible to 10m-class ground-based telescopes.

Due to the extraordinarily high angular resolution afforded by caustic
crossings, fine structures in and around the planet are, in principle,
also detectable.  We first presented a brief discussion on the
existence and stability of satellites, rings and atmospheric features
of close-in planets, concluding that although rings and satellites may
be short-lived due to dynamical forces, the ultimate fate of such
structures is not clear.  There are good reasons to believe that
atmospheric features may be important in close-in planets.  We
therefore considered the signatures of satellites, rings, spots, zonal
bands, and non-uniform surface brightness profiles on the light curves
of planetary caustic-crossings.  Where possible, we used semi-analytic
approximations to derive useful expressions for the magnitude of the
deviations expected for these features, as a function of the relevant
parameters, such as the albedo or size of the feature.  We express
these deviations in terms of $\delta_{\rm p}$, the magnitude of the
planetary deviation.

We find that rings produce deviations of amplitude $\sim 10\%
\delta_{\rm p}$, whereas spots, zonal bands, and satellites all produce
deviations of order $\sim 1\%\delta_{\rm p}$.  These semi-analytic
estimates are supported by more detailed numerical simulations.  
We also find that the light curve of a planet with the surface 
brightness profile expected from Lambert scattering deviates from 
that of a uniform source by $\sim 10\%\delta_{\rm p}$.  This affords 
the possibility of probing the physical processes of the atmospheres
of distant extrasolar planets by constraining their surface brightness 
profiles, and therefore the scattering properties of their
constituent particles. 

We assessed the detectability of spots, rings, satellites, and bands
with current and future telescopes.  We found that, for reasonable
assumptions and 10m-class telescopes, the planetary deviation will
have a signal-to-noise of $\sim 15$, a ring system will only
be marginally detectable with a signal-to-noise of $\sim 6$, and all 
other features will be completely undetectable.   For
30m-class or larger telescopes, rings should be easily detectable,
The detection of the non-uniform nature of the planetary
surface brightness profile arising from Lambert scattering requires 100m-class
telescopes for bare detection.  Spots, satellites and zonal bands are
essentially undetectable for even the largest telescopes apertures.

\smallskip

\acknowledgements
We would like to thank Eugene Chiang, Sara Seager, and Gil Holder
for invaluable discussions.  We would also like to thank the anonymous
referee for useful comments and suggestions. 
This work was supported by the 
Astrophysical Research Center for the Structure and Evolution of the Cosmos 
(ARCSEC) of 
Korean Science and Engineering Foundation (KOSEF) through 
Science Research Center 
(SRC) program, and 
by NASA through a Hubble Fellowship grant
from the Space Telescope Science Institute, which is operated by the
Association of Universities for Research in Astronomy, Inc., under
NASA contract NAS5-26555.

\clearpage

\end{document}